# KINEMATICS OF ELECTRONS IN THE VOLUME OF A PLANAR VACUUM DIODE

**Dimitar G. Stoyanov**

Faculty of Engineering and Pedagogy in Sliven, Technical University of Sofia
59, Bourgasko Shaussee Blvd, 8800 Sliven, BULGARIA
e-mail: dgstoyanov@abv.bg

**Abstract**

*The kinematics laws of electrons motion in the volume of planar vacuum diode are obtained. The physically acceptable initial and boundary conditions for the regime of changing of current are presented. A new solution allowing diode self-clogging is suggested.*

## 1. INTRODUCTION

During the studying of physical processes taking place in ensembles of identical particles a one-particle approximation is possible to be applied in a series of cases. Within this approximation the motion of a single particle in a summary field created by the interaction of the given particle with the other particles of the ensemble is regarded. By applying of this approximation some differential equations describing the kinematics of the particles are often obtained. It turn out that it is necessary the obtained mathematical solutions to be very carefully analyzed and interpreted because some of them have no physical sense.

The goal of this article is the solutions of such differential equation to be analyzed, and a physically well-grounded solution of the motion of electrons in the volume of planar vacuum diode to be selected.

## 2. KINEMATICS OF THE ELECTRON

### 2.1 Theoretical formulation

The planar vacuum diode presents a system of two parallel metal plain electrodes with an area **S**, and situated apart at a distance **d**. One of the electrodes (the cathode) emits electrons with electrical charge **q** (**q** < 0). Within our analysis the initial velocity of electrons emission is equal to $V_0$, and the electrons form a density of the emitted current $j_0$ [1, 3]. The direction of the vector of the initial velocity is perpendicular to the cathode. We assume that the cathode lies in the plane **YOZ** of the Cartesian coordinate system and has a coordinate $x = 0$. Hence the anode will have a coordinate $x = d$. We suppose that $d$ is sufficiently great and it is always bigger than $x_m$. At such geometry of the problem all physical vectors will be in parallel with the axis $OX$, and all magnitudes will depend only on the coordinate $x$.

The density of the electric charge $k$ in each point of the space between both electrodes is defined by the equation of Poison in one-dimensional case [1, 3]:

$$\frac{dE}{dx} = \frac{k}{\varepsilon_0} \tag{1}$$

Where, $E$ is the intensity of the electrical field, and $\varepsilon_0$ is the absolute permittivity in vacuum.

In a stationary regime a current with density $j$ is flowing through the planar vacuum diode. The magnitude of this density is constant in the volume [1, 3], and can be expressed by the electrons velocity $V$ and the density of the electric charge $k$.

$$j = k.V \tag{2}$$

And according to the second principle of Newton

$$m.\frac{dV}{dt} = q.E \tag{3}$$

By substitution of (1) and (3) in (2) we obtain

$$j = \varepsilon_0.V.\frac{d}{dx}\left(\frac{m}{q}.\frac{dV}{dt}\right). \tag{4}$$

Taking into account that $dt = dx/V$ [2] we obtain

$$\frac{d}{dt}\left(\frac{dV}{dt}\right) = \frac{d^3x}{dt^3} = \frac{j.q}{\varepsilon_0.m} = b = \text{const.} \tag{5}$$

This is the differential equation which the coordinate of the electron during its motion in the volume of the planar vacuum diode is obeying on.

**2.2 General solution of the equation**

We will find the general solution $x(t)$ of equation (5) by integration in stages. Time will be read from the moment of electrons emission.

After the first integration of eq. (5) the acceleration of the electron is obtained:

$$a(t) = \frac{d^2x}{dt^2} = b.(t - t_m). \tag{6}$$

In eq. (6) with the term $t_m$ is expressed the time moment in which the acceleration is nullified.

The second integration of eq. (5) gives the velocity of the electron

$$V(t) = \frac{dx}{dt} = V_0 + b \cdot \left[\frac{(t-t_m)^2}{2} - \frac{t_m^2}{2}\right]. \tag{7}$$

If $t = t_m$ the electron will possess minimal velocity $V_m$

$$V_m = V_0 - b \frac{t_m^2}{2}. \tag{8}$$

Taking into account eq. (8) we can reduce eq. (7) in the form

$$V(t) = \frac{dx}{dt} = V_m + b \cdot \frac{(t-t_m)^2}{2}. \tag{9}$$

The third integration of eq. (5) gives the law of electron motion

$$x(t) = V_m \cdot t + b \cdot \left[\frac{(t-t_m)^3}{6} + \frac{t_m^3}{6}\right]. \tag{10}$$

At the moment $t = t_m$ the electron will have the coordinate $x = x_m$

$$x_m = V_m \cdot t_m + b \cdot \frac{t_m^3}{6}. \tag{11}$$

This allows eq. (10) to be represented as

$$x - x_m = V_m \cdot (t - t_m) + b \cdot \frac{(t-t_m)^3}{6}. \tag{12}$$

From the dependence eq. (9) we can obtain:

$$(t - t_m) = \pm \sqrt{\frac{2(V - V_m)}{b}}. \tag{13}$$

If we substitute eq. (13) in (12) we will obtain

$$x - x_m = \pm \frac{\sqrt{2}}{3 \cdot \sqrt{b}} \cdot \left[(V - V_m)^{3/2} + 3 \cdot V_m \cdot (V - V_m)^{1/2}\right] \tag{14}$$

**2.3. An analysis of the solution**

From the point of view of mathematics the obtained equation is a standard one.

The analysis will be done as the magnitudes will be made dimensionless. As a scale for the velocity, for the coordinate $x$, and for the time we will use $V_0$, the distance between the electrodes $d$ and $d/V_0$, respectively.

After making the scale eq. (5) will be of the form:

$$\frac{d^3x}{dt^3} = \frac{j \cdot q}{\varepsilon_0 \cdot m} \cdot \frac{d^2}{V_0^3} = \beta = \text{const}. \tag{15}$$

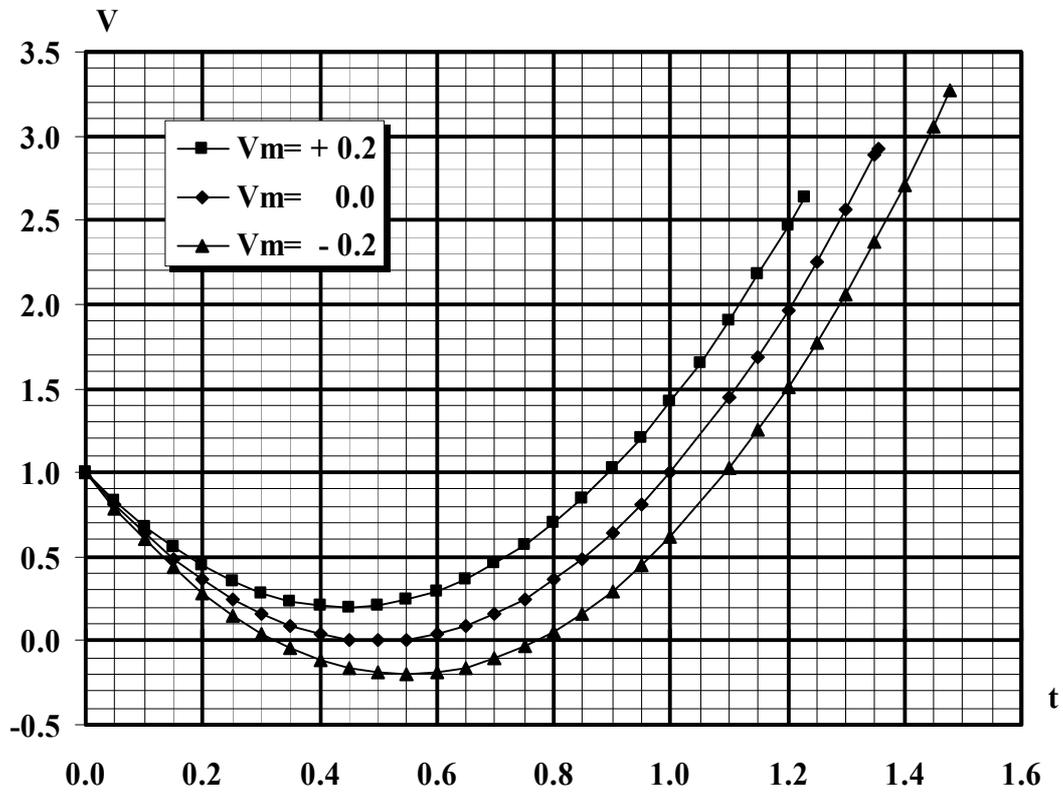

Fig. 1. A graphic dependence of the velocity law at β = 8, initial velocity $V_0$ = 1, and various $V_m$ ( 1 – $V_m$ = +0.2; 2 – $V_m$ = 0.0 and 3 – $V_m$ = -0.2).

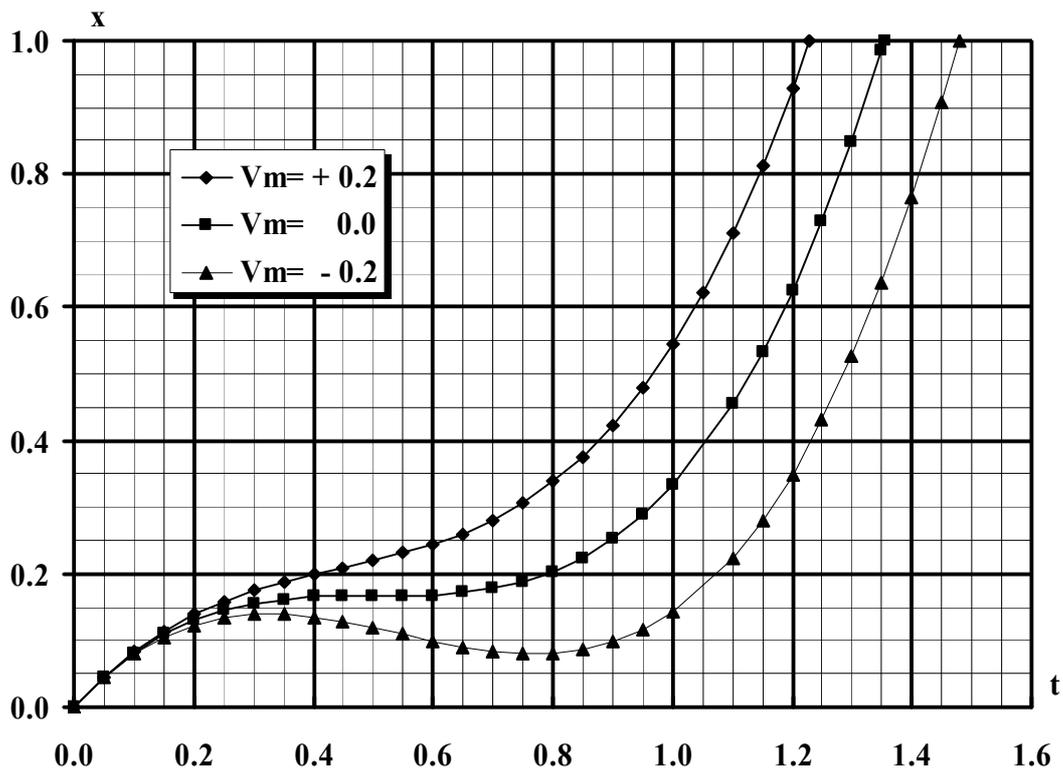

Fig. 2. A graphic dependence of the motion law at β = 8, initial velocity $V_0$ = 1, and various $V_m$ ( 1 – $V_m$ = +0.2; 2 – $V_m$ = 0.0 and 3 – $V_m$ = -0.2).

Fig. 1 and 2 represent the exemplary graphic dependences of eq. (9) and eq. (12) at equal $\beta = 8$, and at identical initial velocity $V_0 = 1$ and different $V_m$ (curves 1 – $V_m$ = +0.2; 2 – $V_m = 0.0$ and 3 – $V_m = -0.2$). These graphic dependences are also standard and likely.

The unexpected contradictions occur when the graph of the coordinate x dependence of the velocity of the particles is built (see Fig. 3).

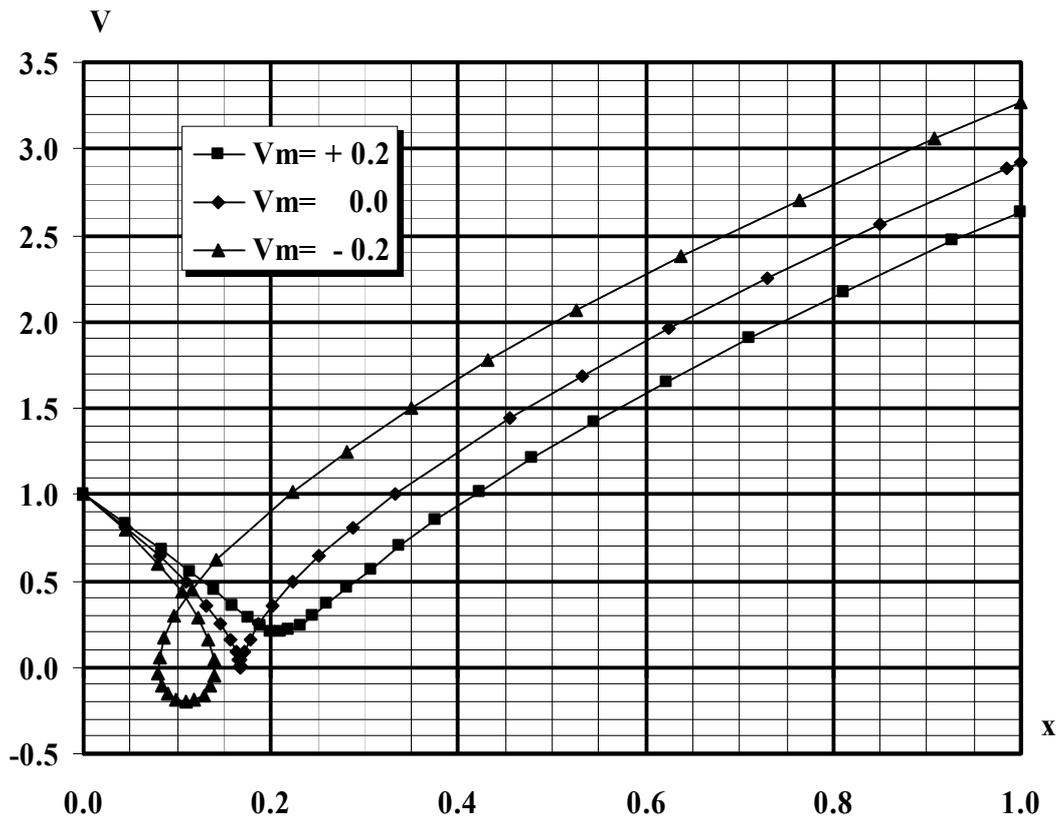

Fig. 3. Coordinate x dependence of particles velocity at $\beta = 8$, and various $V_m$ ( 1 – $V_m$ = +0.2; 2 – $V_m = 0.0$ and 3 – $V_m = -0.2$).

As it is evident on Fig. 3 there is an ambiguity in the velocity dependence on the coordinate **x**. At one and the same magnitude of the coordinate **x** the velocity has up to three different magnitudes in a definite region of curve 3. This mathematical result is in contradiction with the paradigm in physics according to which at a definite point in the space where there are potential fields, upon the charged particle only one unique in magnitude force acts, and there is only one unique in magnitude velocity. Therefore, the solutions of (5) for which $V_m < 0$, are unacceptable from the view point of physics.

This is the reason the finding of another physically based equation to be necessary.

**2.4 A new solution of the problem**

One such solution exists, and it is that it is possible the beam of electrons to begin to get clogged at certain circumstances.

A) If $V_m > 0$ all electrons emitted from the cathode reach to the anode. A current with density $j_0$ will flow through the diode.

At this regime the constant $\beta$ will have the value

$$\beta_0 = \frac{q}{\varepsilon_0.m} \cdot \frac{d^2}{V_0^3} \cdot j_0 \tag{16}$$

Recording eq. (14) twice for the cathode and anode, and after summing up both expressions we obtain

$$\frac{\sqrt{V_0 - V_m}.(V_0 + 2.V_m)}{\sqrt{\beta_0}} + \frac{\sqrt{V_d - V_m}.(V_d + 2.V_m)}{\sqrt{\beta_0}} = \frac{3}{\sqrt{2}}. \tag{17}$$

With the help of eq. (17) we can calculate the velocity of the electron upon the anode $V_d$ at a given magnitude of the velocity in the minimum $V_m$. On the voltage-current characteristic of the vacuum diode this dependence represents a horizontal straight line.

This is **a regime of saturation of current through the vacuum diode**.

B) If $V_m = 0$ it is impossible all electrons emitted from the cathode to reach to the anode. In the sense that when the electrons, emitted from the cathode with current density $j_0$ get into $x_m$, they will move with zero acceleration according to (6), but and with zero velocity $V_m$. This allows a part of them to begin to move back to the cathode.

This part of them which reaches to the anode will form a current density $j$ ($j < j_0$) through the vacuum diode. The rest electrons forming a current density $j_0 - j$ move back to the cathode, and they are absorbed by it. Those electrons moving to the cathode will possess in the given geometric point with coordinate $x \in (0, x_m]$ the same in magnitude velocity as that possessed by the electrons emitted from the cathode which have not yet reached $x_m$. This is due to the fact that the condition, the kinetic energy in this point to be identical for both types of electrons, is fulfilled. The moving back electrons will increase the density of the electric charge in this region.

This is a regime of **changing of the current passing through the vacuum diode.**

This leads to pre-determination of the differential equation (5). In the regions from the cathode to the minimum, and from the minimum to the anode the constant $\beta$ from eq. (5) will be different:

$$\beta_1 = \frac{q}{\varepsilon_0 . m} \cdot \frac{d^2}{V_0^3} (2.j_0 - j) \qquad x < x_m \qquad (18a)$$

$$\beta_2 = \frac{q}{\varepsilon_0 . m} \cdot \frac{d^2}{V_0^3} j \qquad x > x_m \qquad (18b)$$

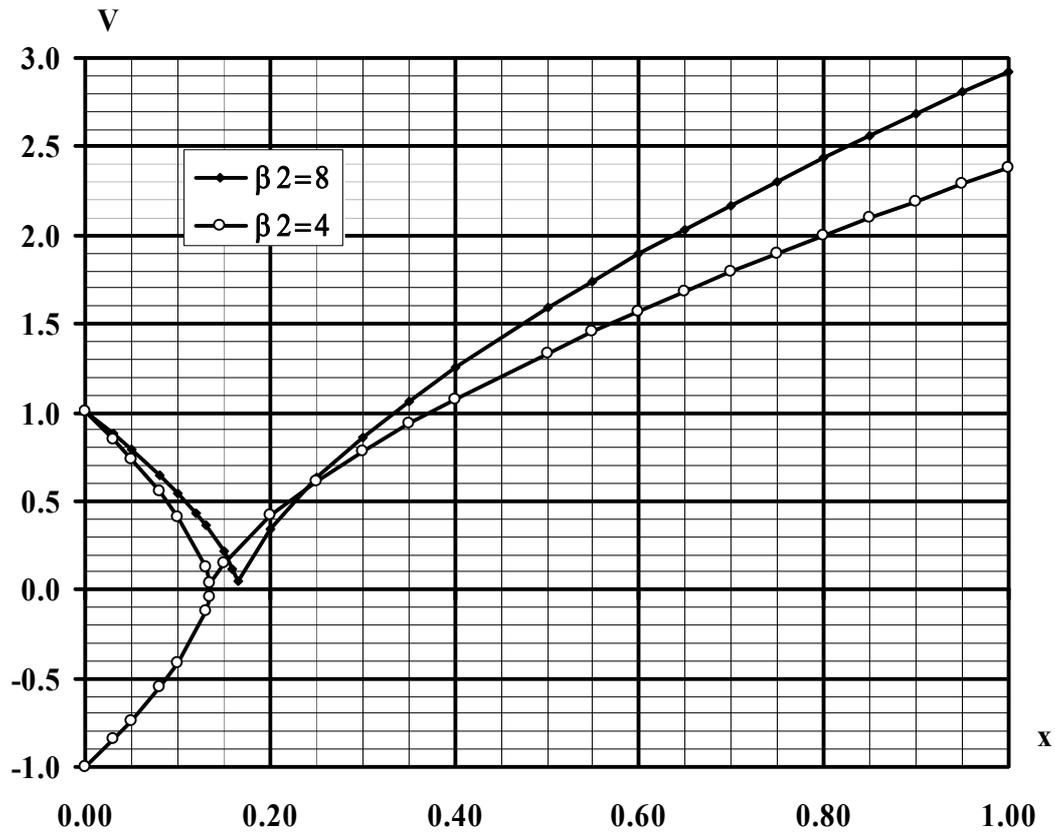

Fig. 4. Coordinate x dependence of the velocity of electrons according to (19a) and (19b) at: 1 - $\beta_1 = 8$, $\beta_2 = 8$; 2 - $\beta_1 = 12$, $\beta_2 = 4$.

For that reason, for the velocities from eq. (14) we will obtain the following dependences

$$V^{3/2} = \frac{3 \cdot \sqrt{\beta_1}}{\sqrt{2}} \cdot (x_m - x) \qquad x < x_m \qquad (19a)$$

$$V^{3/2} = \frac{3 \cdot \sqrt{\beta_2}}{\sqrt{2}} \cdot (x - x_m) \qquad x > x_m \qquad (19b)$$

The position of the minimum is determined from eq. (19a) recording the expression for the cathode

$$V_0^{3/2} = \frac{3 \cdot \sqrt{\beta_1}}{\sqrt{2}} \cdot x_m \qquad x = 0 \qquad (20)$$

In Fig. 4 the coordinates dependence of the velocity of electrons according to eq. (19a) and (19b) for both sets of values of $\beta_1$ and $\beta_2$ is presented.

The velocity of the electrons upon the anode $V_d$ can be obtained from eq. (19b)

$$V_d^{3/2} = \frac{3 \cdot \sqrt{\beta_2}}{\sqrt{2}} \cdot (1 - x_m) \qquad x = 1 \qquad (21)$$

Combining eq. (20) and eq. (21) we obtain

$$\frac{V_0^{3/2}}{\sqrt{\beta_1}} + \frac{V_d^{3/2}}{\sqrt{\beta_2}} = \frac{3}{\sqrt{2}}. \qquad (22)$$

Thus the obtained expression eq. (22) represents in hidden form the voltage-current characteristic of a planar vacuum diode in a regime of changing of the current.

Actually, if we accept that the voltage of the anode in respect to the cathode in dimensionless form, with scale for the voltage $m \cdot V_0^2 /(2 \cdot q)$ is

$$U = V_d^2 - 1. \qquad (23)$$

In order the correspondence with [1, 3] to be available let us put that the density of the current through the diode in dimensionless units $J$ is

$$J = 2 \cdot \beta_2. \qquad (24)$$

If we substitute eq. (23) and eq. (24) in eq. (22) we will obtain

$$\frac{1}{\sqrt{2 \cdot J_0 - J}} + \frac{(U+1)^{3/4}}{\sqrt{J}} = \frac{3}{2}. \qquad (25)$$

It represents a voltage-current characteristic of a vacuum diode in dimensionless units for the regime of changing of the current through the diode [3].

Using (22), (23) and (24) for the set of values $\beta_0 = 8$, $\beta_1 = 2\beta_0 - \beta_2$, and $\beta_2 \in (0, 8]$ as an illustration we can calculate $V_d$ and build Fig. 5 ($J_0 = 2\beta_0 = 16$).

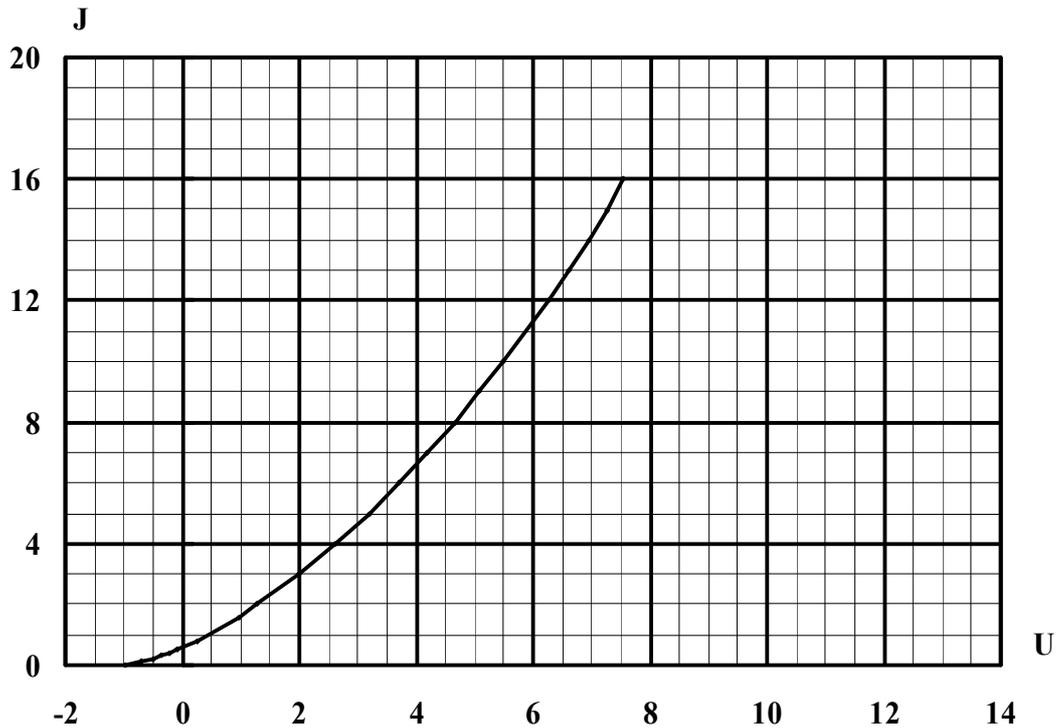

**Fig. 5. A voltage-current characteristic of a planar vacuum diode in dimensionless units.**

## 3. CONCLUSION

The kinematics laws for the motion of electrons in the volume of a planar vacuum diode are obtained. The physically acceptable initial and boundary conditions for a regime of changing of the current are pointed out.

**REFERENCES**

[1] Stoyanov, D. G., "A correct formulation of Langmuir problem", *Announcements of Union of Scientists – Sliven,* v.3, pp 43-45 (2001) (in Bulgarian).

[2] Stoyanov, D. G., "A kinematics formulation of Langmuir problem", *Announcements of Union of Scientists – Sliven,* v.3, pp 46-48 (2001) (in Bulgarian).

[3] Stoyanov D. G., "Planar vacuum diode with monoenergetic electrons", *J. Appl. Electromagnetism*, V 8, No1, 2, June 2006, December 2006, pp 35-48 (2006)